\documentclass[
aps,
prb,
twocolumn,
superscriptaddress,%
showpacs,%
preprintnumbers,%
byrevtex,%
floatfix%
]{revtex4-2}

\usepackage{dcolumn}
\usepackage{bm}
\usepackage{ifthen}
\usepackage{xcolor}
\usepackage{amssymb}
\usepackage{textcomp}
\usepackage{amsfonts}
\usepackage{amsmath}
\usepackage[dvips]{graphicx}
\usepackage{graphpap}

\usepackage{multirow}
\usepackage{here}
\usepackage{romannum}

\begin{document}

\date{\today}
\title{Stability and electronic structure of NV centers at dislocation cores in diamond}

\author{Reyhaneh Ghassemizadeh}
\email{reyhaneh.ghassemizadeh@iwm.fraunhofer.de}
\affiliation{Fraunhofer Institute for Mechanics of Materials IWM, W\"ohlerstra{\ss}e 
11, 79108 Freiburg, Germany}

\author{Wolfgang K\"orner}
\affiliation{Fraunhofer Institute for Mechanics of Materials IWM, W\"ohlerstra{\ss}e
	11, 79108 Freiburg, Germany}

\author{Daniel F. Urban}
\affiliation{Fraunhofer Institute for Mechanics of Materials IWM, W\"ohlerstra{\ss}e 11, 79108 Freiburg, Germany}

\author{Christian Els\"asser}
\affiliation{Fraunhofer Institute for Mechanics of Materials IWM,
W\"ohlerstra{\ss}e 11, 79108 Freiburg, Germany}%
\affiliation{University of Freiburg, Freiburg Materials Research Center (FMF), Stefan-Meier-Stra{\ss}e 21, 79104 Freiburg, Germany}

\begin{abstract}
We present a density functional theory analysis of the negatively charged nitrogen-vacancy (NV) defect 
complex located at or close to the core of $30^{\circ}$ and $90^{\circ}$ partial glide dislocations in diamond.  
Formation energies, electronic densities of states, structural deformations, hyperfine structure and zero-field 
splitting parameters of NV centers in such structurally distorted environments are analyzed. 
The formation energies of the NV centers are up to $3$ eV lower at the dislocation cores compared to the bulk values of crystalline diamond. 
We found that the lowest energy configuration of the NV center at the core of a $30^{\circ}$ partial glide dislocation is realized when the axis of the NV center is oriented parallel to the dislocation line. This special configuration has a stable triplet ground state. Its hyperfine constants and zero field splitting parameters deviate by only $3\%$ from values of the bulk NV center. Hence, this is an interesting candidate for a self-assembly of a linear array of NV centers along the dislocation line.
\end{abstract}


\maketitle
\section{Introduction}

\begin{figure}[]
\setlength{\unitlength}{1mm}
\begin{picture}(80,41)(0,0)
\put(-5,0){\includegraphics[width=1.\columnwidth]{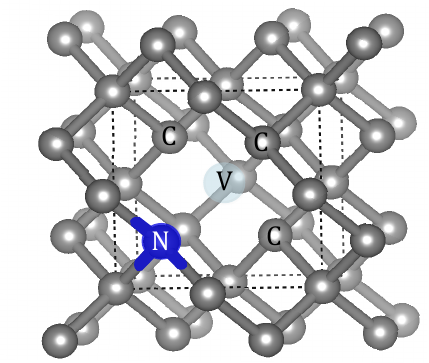}}
\put(42,0){\includegraphics[width=1.\columnwidth]{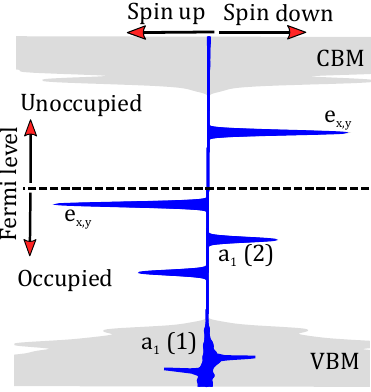}}
\put(0,40){(a)}
\put(40,42){(b)}  
\end{picture}
\caption{(a) Atomic structure of a NV center in diamond. The N atom is drawn as a dark blue sphere and the C vacancy (V) is highlighted with a light blue sphere. The three C atoms next to the vacancy are labeled, the further C atoms are depicted in gray.  (b) Schematic plot of electronic energy levels of NV$^-$ defect complex in the $S=1; m_s=1$ triplet ground state (blue). The electronic density of states of the diamond host crystal is plotted in gray. Two electrons occupy the $a_{1}(1)$ levels below the valence band maximum (VBM) of diamond. Two electrons occupy the $a_{1}(2)$ levels and two electrons form the triplet of the $e_{x,y}$ states in the band gap of diamond.}\label{fig:nv}
\end{figure}

The negatively charged nitrogen-vacancy (NV) center in diamond is a point-defect complex that consists of a substitutional nitrogen 
atom in the diamond crystal structure, a vacant carbon site next to it and an additional electron (see Fig.~\ref{fig:nv}(a)). NV centers are well known for their remarkable optical and magnetic properties and their long spin-coherency time, which make them excellent candidates for the use as sensors in spatial-atomic-resolution quantum magnetometry~\cite{ba08,ma08,ac09} and as qubits in solid-state-based quantum computing~\cite{jac09,wolf21, pez21}.
The six valence electrons of the NV center are localized at the defect complex. Four of them contribute to electronic defect levels lying within the band gap of diamond (see Fig.~\ref{fig:nv}
(b)), which are then accessible by single-photon absorption in the visible light range of the electromagnetic 
spectrum. Moreover, the electronic configuration of the ground state is a S=1 spin triplet state with the three levels marked by $m_{s} 
=0$ and $m_{s}= \pm{1}$. Their degeneracy in the absence of magnetic field and at room temperature is lifted by $2.87$ GHz~\cite{fel09}, 
which is known as zero-field splitting (ZFS). While the ZFS originates from the self-interaction of electron spins of the system, 
the interaction of electron spins with nuclear spins in their environment (spins of $^{13}C$ and $^{14}N$ nuclei),
promote the hyperfine interaction, which influences the spin coherency of the NV centers. 

Despite many efforts to synthesize purified diamond with lowest possible density of paramagnetic impurities
still crystallographic defects such as low-angle grain boundaries, stacking faults, or  
dislocations emerge during the growth process of diamond~\cite{blu03,ach20}. These extended defects are immobile even at the high annealing temperature, where diffusion of vacancies supports the formation of NV centers~\cite{ach20,will06}. High densities of dislocations are reported\cite{ach20,will06,jon09,schreck16} for both techniques that are currently used to fabricate diamonds with optimized properties for industrial and high-tech applications, namely the high-pressure high-temperature (HPHT) and chemical vapor deposition (CVD) techniques. Therefore, it seems highly relevant to study 
the influence of such extended crystallographic defects on the physical properties of a NV center that are crucial for their applicability in quantum 
technology, in particular its stability, ZFS and hyperfine structure (HFS) parameters. The influence of some of these defects on NV center properties have been recently studied~\cite{ko21,ko22,loef22}.

In this paper we study the influence of dislocation cores on the properties of NV centers. Dislocations are common 
defects in diamond imposing a long-range translational symmetry breaking of the perfect single crystal along the so-called dislocation line. 
An interesting perspective is whether the presence of a dislocation offers the possibility to obtain a linear chain of NV centers by some self-assembly mechanism along the dislocation line and how the electromagnetic properties of such aligned NV centers are modified with respect to the perfect crystal. 
This is of interest for applications since in quantum magnetometry the signal can be enhanced by using not only a single NV center but coupled
NV centers. Such an ensemble has an enhanced magnetic sensitivity which scales with $\sqrt{X}$ due to the higher 
fluorescence of X sensing NV centers~\cite{st10}.
Furthermore, hundreds or better thousands of coupled NV centers 
qubits will be needed to build quantum registers for practically useful solid-state based quantum computers in the future.
Here, the controlled patterning of NV center qubits remains an open challenge. Recently, the decoration of NV centers at the specific growth sector boundary of \{111\}-\{113\} surface facet on diamond single crystal with the decoration width of $\approx 5 \mu m$ has been experimentally observed~\cite{dig20}, in which no preferential NV center orientation was detected. However, for getting a chain of aligned NV centers we propose the investigation of NV center decoration of linear crystal defects, namely dislocation cores.

For our theoretical study the formulated perspective translates into the following questions: 
(i) Are there energetically preferred sites for NV centers at or near the cores of dislocations?
If the answer is yes:
(ii) How does the electronic level structure at those sites look like? 

In order to answer these questions, we use a density functional theory (DFT) analysis of atomistic supercell models containing both, dislocation cores and NV centers.   
We analyze the defect formation energy, structural deformations, electronic defect levels, HFS and ZFS parameters in order to quantify the influence of dislocations on NV center defect levels at the their core or in their vicinity.

The manuscript is organized as follows: In Sec.~\ref{sec:theory} 
the details of the atomistic supercell models and the DFT calculations are described.
The results for formation energies of NV$^-$ defect complexes, electronic densities of states (DOS) and structural relaxations are presented in Secs.~\ref{sec:results:energy}, \ref{sec:results:elec}, and \ref{sec:results:geo}, respectively. 
The results for the ZFS are reported and discussed in Sec.~\ref{sec:results:zfs}, and the analysis of the HFS is 
presented in Sec.~\ref{sec:results:hyperfine}. 
Section ~\ref{sec:summary} summarizes our findings.

\begin{figure*}[]
	\setlength{\unitlength}{1mm}
	\begin{picture}(170,100)(0,0)
	\put(-4,56){\includegraphics[width=1.\columnwidth]{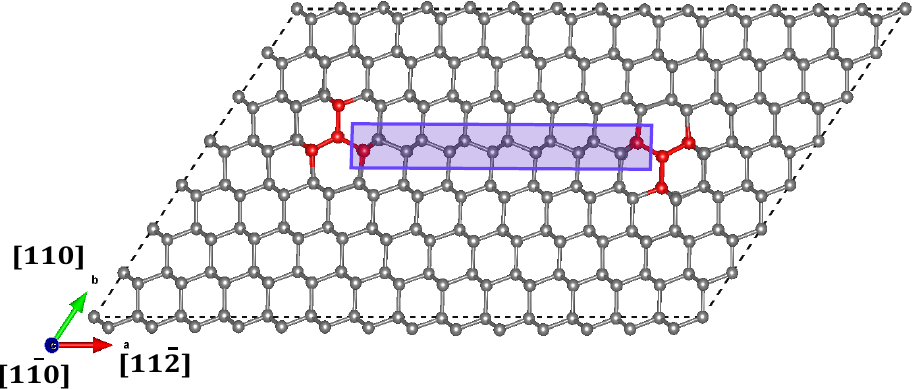}}
	\put(92,58){\includegraphics[width=1.\columnwidth]{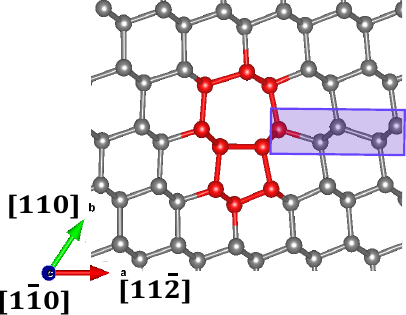}}
	\put(135,58){\includegraphics[width=1.\columnwidth]{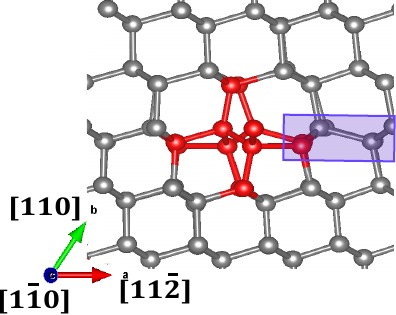}}
	\put(-4,0){\includegraphics[width=1.\columnwidth]{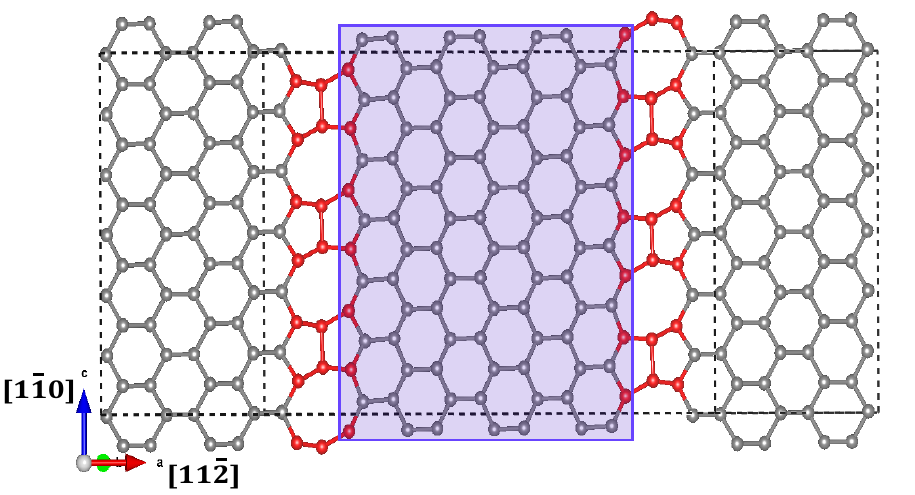}}
	\put(92,10){\includegraphics[width=1.\columnwidth]{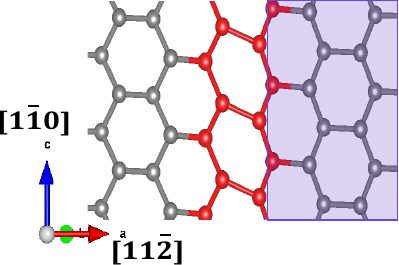}}
	\put(135,10){\includegraphics[width=1.\columnwidth]{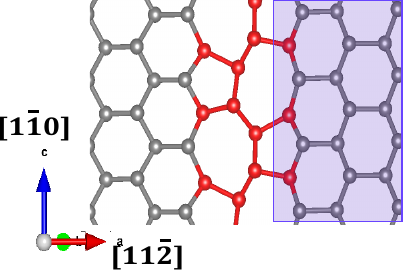}}
	
	\put(35,100){\textbf{$30^\circ$ Partial glide}}
	\put(97,100){\textbf{$90^\circ$ SP Partial glide}}  
	\put(140,100){\textbf{$90^\circ$ DP Partial glide}}
	
	\put(35,52){(a)}
	\put(112,52){(b)}  
	\put(158,52){(c)}
	\put(35,-2){(d)}  
	\put(112,-2){(e)}
	\put(158,-2){(f)}  
	
	\end{picture}
\caption{(a) Atomistic supercell model of the diamond crystal including two oppositely oriented $30^\circ$ partial glide dislocation cores (in red) after reconstruction and an intrinsic stacking fault (shaded in purple) between the two cores. The dashed lines indicate the supercell size and shape. (b) $90^\circ$ SP and (c) $90^\circ$ DP partial glide dislocation cores in the same supercell (to save space only a zoom-in to one of the cores is displayed). The upper panels are views in the direction of the dislocation lines and the bottom panels (d),(e) and (f) are perpendicular views onto the dislocation lines.}\label{fig:30_deg}
\end{figure*}
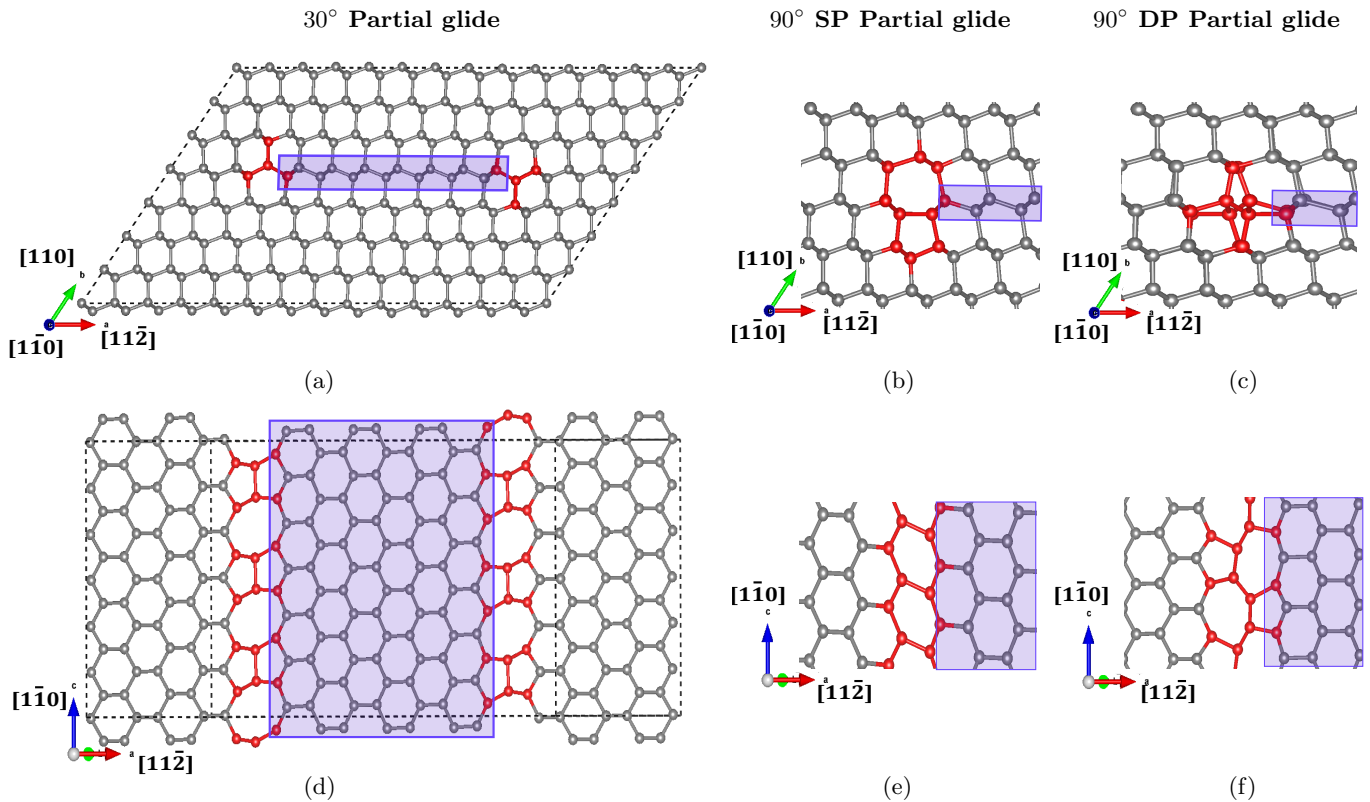

\section{Theoretical approach}\label{sec:theory}

\subsection{Atomistic supercell models}\label{sec:supercell}

\subsubsection{Construction of dislocation cores}
Dislocations and their cores in diamond-type crystal structures are already discussed in detail in many publications~\cite{bol70,hor58,par15,blu02,blu03,zib15}. Therefore, we here only briefly describe the dislocation core structures used for the purpose of 
this study. We consider the cores of the two most commonly occurring dislocation types in diamond, the $30^\circ$ and $90^\circ$ partial 
glide dislocations~\cite{,heg2000,hor58,blu03,blu02}. The formation of two partial dislocations, which are created from dissociation of one perfect dislocation~\cite{bol70,hor58,blu03}, results in the formation of an intrinsic stacking fault separating the two partials.

Like dislocations in a fcc crystal structure, the dislocations in cubic diamond are typically aligned on
$\{$111$\}$ planes along $<$110$>$ directions. Accordingly, for our choice of dislocation cores the dislocation line is along 
$[1\bar{1}0]$ and the Burgers vectors of the $30^\circ$ and $90^\circ$ partial glide dislocations are respectively chosen to 
be ${b}_{30^\circ} ={1\over6} [2\bar{1}\bar{1}]$ and ${b}_{90^\circ} = {1\over6}[11\bar{2}]$. 
The $90^\circ$ partial glide dislocations have two distinct core reconstructions, namely a single period (SP) and a double period (DP) core~\cite{par15,ewe01,blu02,rod17}. 
In Fig.~\ref{fig:30_deg} the reconstructed cores of the $30^\circ$ and $90^\circ$ SP and DP partial glide dislocations 
in diamond structure are displayed. In the following we refer to them using the abbreviations $30^\circ$, $90^\circ$ SP 
or DP. 

Dislocations break the translational symmetry of the crystal and are accompanied by a long-range elastic strain field.
In order to build a finite size structure model that obeys periodic boundary conditions, the following trick can be used.
By placing two oppositely oriented dislocations of the same kind in one non-tetragonal supercell~\cite{big92} one can form an approximate quadrupole arrangement of dislocations that is repeated periodically. In such a quadruple arrangement the long-range strain fields of the partial dislocations compensate each others to approximately zero in some regions between the dislocation cores. We use a diamond supercell of 
$7a_l\hat{x}\times7a_l\hat{y}\times6 a_l\hat{z}$ with the $a_l (\hat{x},\hat{y},\hat{z})$ being the lattice vectors pointing in the $[11\bar{2}]$, $[110]$ and $[1\bar{1}0]$ crystal direction respectively, and
$a_l$ being the lattice constant of 3.567 $\rm{\AA}$ taken from experiment ~\cite{hol91}. This supercell contains 1176 carbon atoms and is at the size-limit concerning our available computational resources. The separation of the two dislocation cores along the stacking fault that connects them is $\sim 10 \rm{\AA}$. 

Furthermore, it is important to ensure large enough supercell dimensions in order to avoid the electrostatic self-interaction of the defect with its periodic images. In case of the NV center a minimum distance of 15 $\rm{\AA}$ is needed to avoid the covalent  bond hybridization effects of neighboring NV states. The dimension of our chosen supercell is 30.5 $\rm{\AA}$, 17.6 $\rm{\AA}$ and 15.1 $\rm{\AA}$ in $x$, $y$ and $z$ direction, respectively. 

All the atomistic structures in this study were created using the atomistic simulation environment (ASE) 
package~\cite{ase} and the dislocation cores were built with the ASAP3 package 
implemented within ASE. 

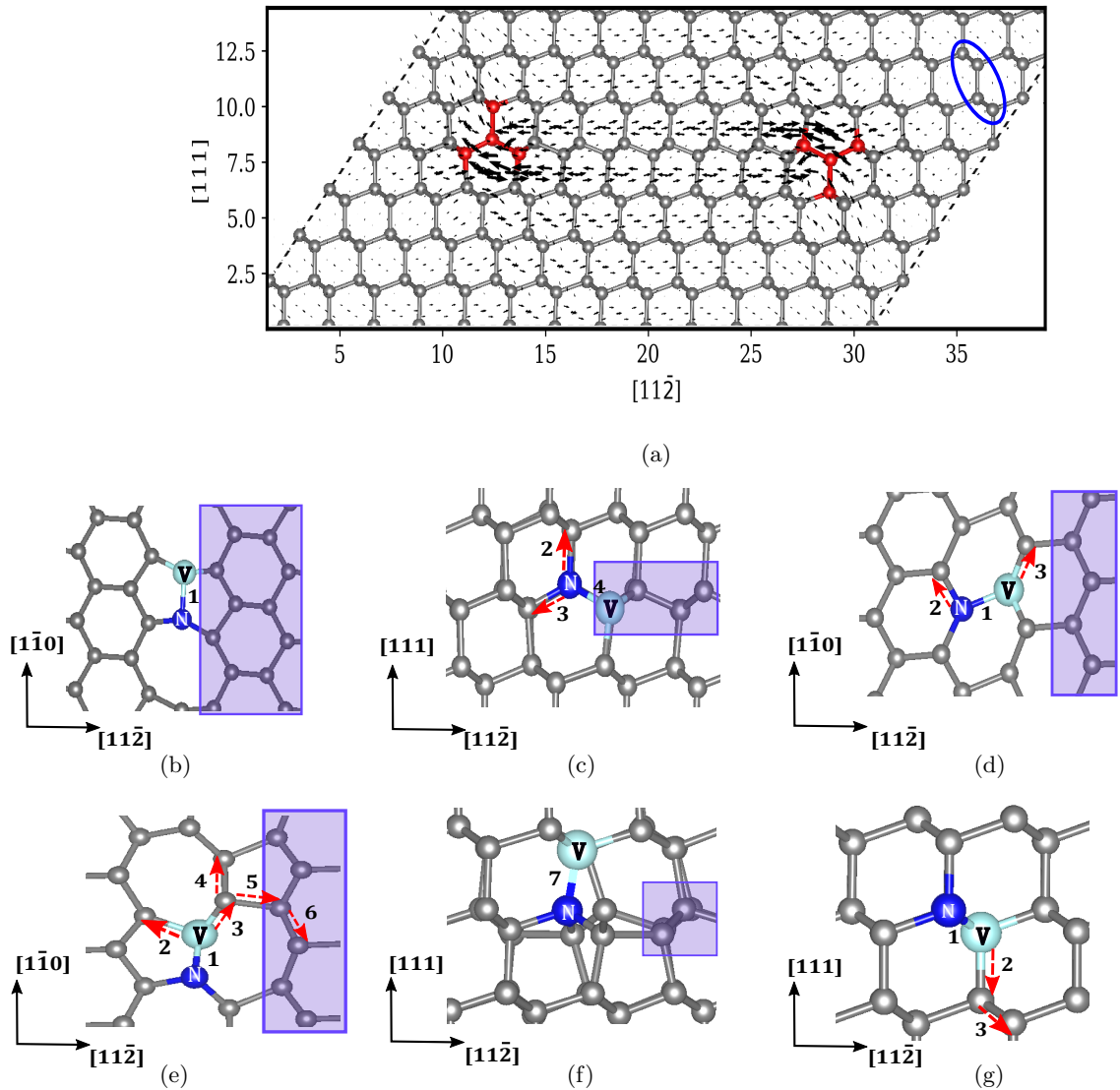
\begin{figure*}[]
	\begin{center}
		\setlength{\unitlength}{1mm}
		\begin{picture}(170,140)(0,0)
		\put(25,85){\includegraphics[width=1.2\columnwidth]{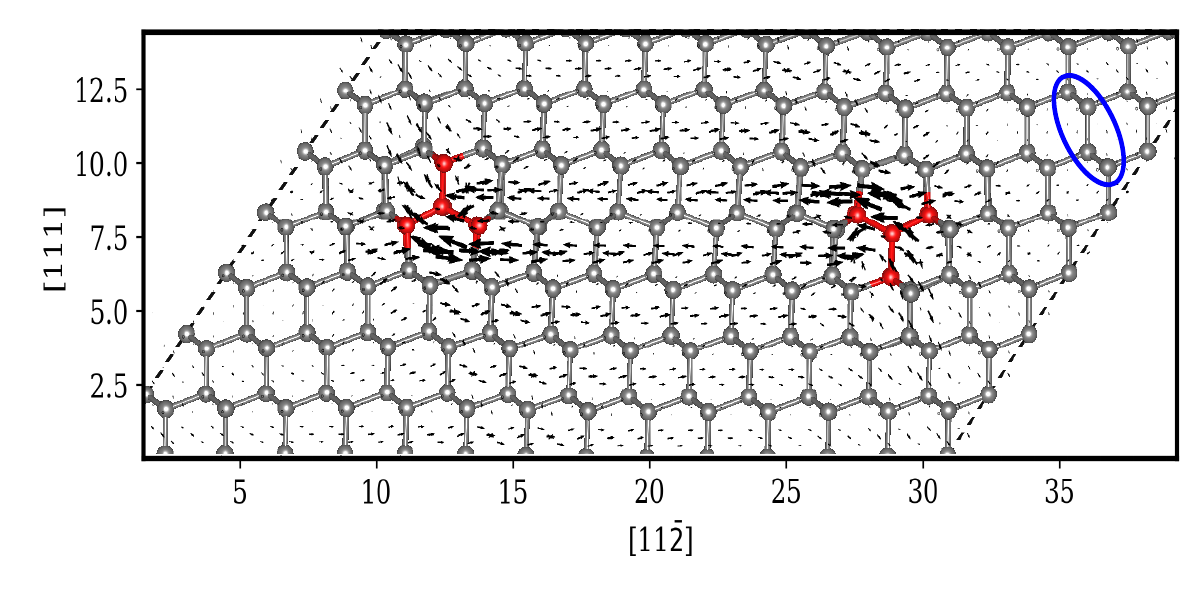}}
		\put(5,43){\includegraphics[width=1.\columnwidth]{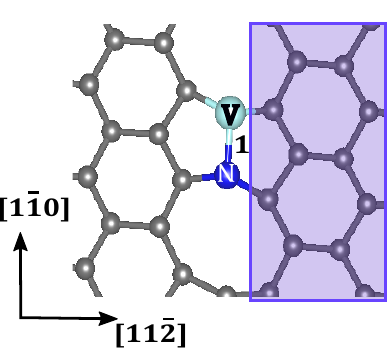}}
		\put(55,43){\includegraphics[width=1.\columnwidth]{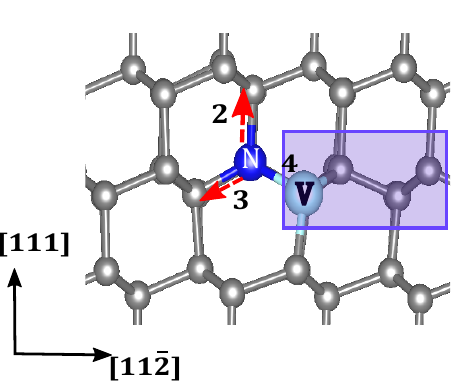}}
		\put(110,43){\includegraphics[width=1.\columnwidth]{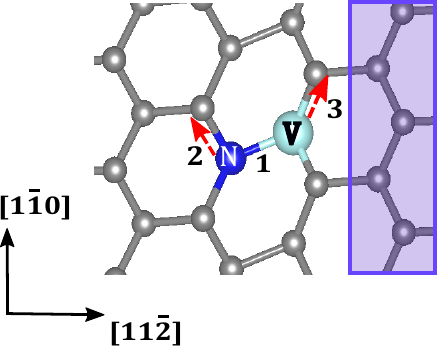}}
		\put(5,0){\includegraphics[width=1.\columnwidth]{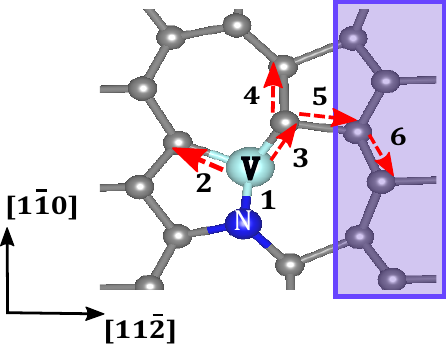}}
		\put(55,0){\includegraphics[width=1.\columnwidth]{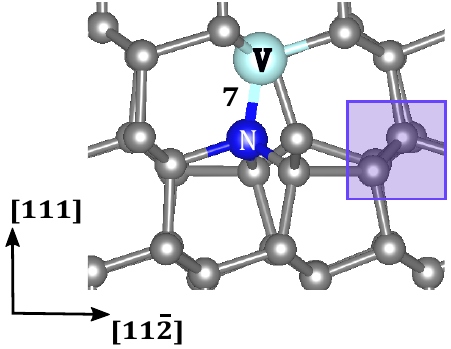}}
		\put(110,0){\includegraphics[width=1.\columnwidth]{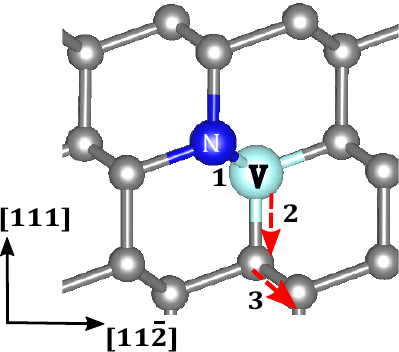}}
		
		\put(90,82){(a)}
		\put(25,40){(b)}  
		\put(80,40){(c)}
		\put(135,40){(d)}  
		\put(25,-2){(e)}
		\put(80,-2){(f)}  
		\put(135,-2){(g)}   
		
		\end{picture}
	\end{center}
	\caption{The upper panel (a) is a differential displacement map (DDM) of the $30^\circ$ dislocation core. 
	Core positions are highlighted in red and the quasi-bulk region for placing the NV defect 
    is indicated by the blue ellipse.
	    The lower panels are zoom-in views of possible NV defect positions at (b) and (c) $30^\circ$, (d) 
		$90^\circ$ SP, (e) and (f) $90^\circ$ DP dislocation cores and (g) the quasi-bulk sites. The latter are the same regardless of the 
		type of the dislocation cores in the supercell. Possible positions of NV defect are indicated by numbers 1, 2, 3, etc.; the red 
		arrows are always pointing from N to V (vacancy). Gray spheres indicate C atoms, the light blue sphere indicates the V site and 
		the blue sphere marks the N atom. 
			}
	\label{fig:dis-nv}
\end{figure*}

\subsubsection{NV defect positions at or near dislocation cores }
In order to study the influence of dislocations on NV centers, we have placed this defect complex at various positions of the dislocation cores. The positions can be summarized in two sets.
(i) the extreme regime, where the NV defect is placed at the structurally most disturbed positions, meaning at or around the dislocation core. We call those positions ``core positions". 
(ii) where the strain field of the quadrupole arrangement of dislocations is minimal and therefore, the NV defect is placed in a bulk-like neighborhood. Those positions are denoted as ``quasi-bulk positions". 

The quasi-bulk positions are identified by means of the differential displacement 
map (DDM) analysis, proposed by Vitek et al.~\cite{vit70}, in which the strain field of a dislocation can be visualized (see Fig.~\ref{fig:dis-nv}(a)). It is based on the comparison of the relative position vectors between a defect-free system and a defect system for all pairs of neighbor atoms. 
In Fig.~\ref{fig:dis-nv}(a) the DDM for the $30^\circ$ dislocation core is shown. 
The length of the arrows indicate the relative displacement of the atomic position between the defect-free system and the defect-containing system. The quasi-bulk positions of NV defect are chosen where the strain field and thus the 
lengths of the arrows are minimal.

Figure \ref{fig:dis-nv}(b)-(f) displays some of the investigated NV defect positions at the cores of the $30^\circ$ and $90^\circ$ SP and DP dislocations. 
We assign numbers to the different NV defect configurations.
For each position there are two possibilities for the NV orientation. The 
numbers $1$, $2$, $3$, etc. belong to red arrows pointing from a N atom to a vacancy. The corresponding reversed orientations of N and V are then labeled by $1_i$, $2_i$, etc.. 
All the core configurations which lie in the glide plane of the dislocation are investigated.
However, from the resulting defect formation energies for the core positions $2$, $2_i$ in $30^\circ$ and $7$, $7_i$ in $90^\circ$ DP dislocations when the majority of the NV defect atoms lie above or below the glide plane of the dislocation, the defect formation energy increases to a great 
extent, which means that not all of the NV defect configurations around or at the dislocation cores are favorable with respect to quasi-bulk configurations. Therefore we did not further investigate those unfavorable positions, specifically for the $90^\circ$ SP dislocation. 
Based on the DDM, the regions of quasi-bulk positions are similar 
in all our studied supercell models, regardless of the type of the dislocation core. Figure~\ref{fig:dis-nv}(g) displays the region of quasi-bulk sites in our unrelaxed $30^\circ$ supercell with NV defect placed at positions $1$, $2$, $2_i$, $3$ and $3_i$.

\subsection{Computational details}\label{sec:computation}

For the structural relaxation of the atomistic supercell models and
the calculation of the physical parameters of interest we use the Vienna Ab Initio Simulation Package 
(VASP)~\cite{kr96,kr99}. The Bloch waves of the valence electrons are expanded in a plane-waves basis (with a cutoff 
energy of 420 eV) and the interactions of the valence electrons with the ionic cores are included by projector-augmented-waves (PAW) potentials~\cite{bl94}. 
The exchange-correlation (XC) energy and potential are treated 
in the generalized gradient approximation (GGA), as given by Perdew, Burke and Ernzerhof (PBE)~\cite{Perdew1996}.      
For all the supercells the Brillouin-zone integrals are evaluated using a mesh of $1\times1\times2$ $k$-points
with a Gaussian broadening of the energy levels by 0.05~eV. The positions of the atoms in the supercell with constant volume 
were relaxed until the residual forces acting on them were less than 0.002~eV/~\rm{\AA}\ and the total-energy difference between two 
consecutive ionic relaxation steps was less than $10^{-5}$ eV. 

We investigate the influence of the proximity of the dislocations on the energetic stability of the NV center by a comparison of total energies for a given electronic configuration.
The NV center in its ground state is a S=1 spin system, meaning that it forms a stable triplet electronic configuration. For the NV center in a perfect crystal environment the triplet ground state configuration is always obtained by VASP without any further constrains.  However, in a structurally distorted environment  the stability of the triplet configuration is not granted. We obtained defect formation energies for the singlet and triplet electronic configurations by constraining the difference between all spin-up and all spin-down electrons in the calculations to either 1 (triplet) or 0 (singlet), respectively.  The calculation formula for the defect formation energy at position $p$ for the electronic configuration $q$, which can be either a singlet or a triplet, reads
\begin{equation}
\label{eq:Eform}
E_q(p)=E_{tot,q}^{\rm NV in Disl}-E_{tot}^{\rm Disl}-(E_{tot}^{\rm NV in Bulk}-E_{tot}^{\rm Bulk})
\end{equation}
where $E_{tot,q}^{\rm NV in Disl}$ is the total energy of the supercell containing the NV$^-$ defect in the electronic configuration $q$ at the position $p$ at or near one of the 
two dislocation cores and $E_{tot}^{\rm Disl}$ is the total energy of the supercell containing only the two dislocation cores. The last two terms 
stem from the bulk diamond supercell of the same shape and size with or without a NV$^-$ defect, $E_{tot}^{\rm NV in Bulk}$ 
and $E_{tot}^{\rm Bulk}$. Note that the lowest total energy of the NV center in the bulk crystal is always obtained for its triplet electronic configuration.

We use the density of states (DOS) analysis to determine the defect electronic states within the band gap of diamond. In order to achieve a higher accuracy 
one has to go beyond the PBE-GGA of DFT, which systematically underestimates the band gap of diamond by more 
than 1~eV (GGA value 4.1 eV versus experimental value 5.47eV~\cite{wort08}).
On the other hand, since we need large supercells for an adequate modeling of the dislocation cores in combination with the 
NV$^-$ defect, a use of GW methods or hybrid functionals would exceed our computational resources.
An efficient approach, which is computationally not more expensive than the PBE-GGA, is the DFT-1/2 method  
developed by Ferreira et al.~\cite{fer08}. Although one has to use this approach with caution~\cite{dou19} it has been shown 
by Lucatto et al.~\cite{luc17} that for the NV$^-$ defect in bulk diamond one can obtain the optical transition levels as energy differences of the Kohn-Sham orbitals with high precision. 

In the DFT-1/2 method, one removes half an electron from the last occupied state of the valence
band and adds it to the first unoccupied state of the conduction band. The adjusted potential for such 
calculations can be generated by the online platform developed by Yuan et al.~\cite{yuan}.  For the single crystal of diamond it
has been shown by Xue et al.~\cite{xue} that removal/addition of half an electron is the best choice to obtain the right band gap. We initially optimized the cutoff radius for pure bulk C atoms by maximizing the diamond band gap. A maximum value of the diamond band gap of $5.75$ eV was obtained for a cutoff radius of $2.4$~bohr. Inserting the NV center into the diamond crystal, the choice of the cutoff radius for defect atoms, meaning the three C atoms and the N atom next to the vacancy, is optimized with respect to either absorption or emission energy as explained in detail in Ref.~\cite{fer08}. 
We obtain the optimized values of the transition energies for cutoff radii $2.9$~bohr for the N atom and $2.4$~bohr for the three C atoms next to V (i.e., the same radius as for bulk C atoms). For a better energy resolution, we evaluate the DOS using a finer mesh of $2\times3\times4$ $k$-points.
 
The computation of the HFS tensor components and the ZFS tensor components is performed using subroutines implemented in VASP and including only the $\Gamma$-point due to the high computational costs. The accuracy of such settings and approximations is discussed in our previous work \cite{ko21,ko22}. 
The values $
\gamma{(^{13}{\rm C})}/2\pi$  = 10.7084 MHz/T and 
$\gamma{(^{14}{\rm N})}/2\pi$  = 3.077 MHz/T~\cite{ber04} are used to obtain the hyperfine tensor elements in this 
work.

\section{Results and discussion}\label{sec:results}

\subsection{Formation energies of NV centers}\label{sec:results:energy}

Creation of NV centers is typically happening in two steps: first the substitutional nitrogen atom is created via irradiation process, then the mobile carbon vacancies diffuse during the annealing process \cite{ber17}. For a possible formation of NV center at a dislocation core, it is useful to first check whether the formation of single N$_C$ and V defects at the dislocation core is also energetically preferential. Respectively, we obtain an energy release of about -3.5 eV and -2.7 eV for the formation of a N$_C$ and a V at the core of the 30$^\circ$  dislocation in diamond. Moreover, binding of these two point defects at the dislocation core releases $\approx 2$ eV energy. 
In the following we present the results for the defect formation energy of the NV$^-$ defect complex at every of the previously described positions and we determine whether the triplet or singlet electronic configuration is more stable.

\begin{figure}
{\includegraphics[width=1.\columnwidth]{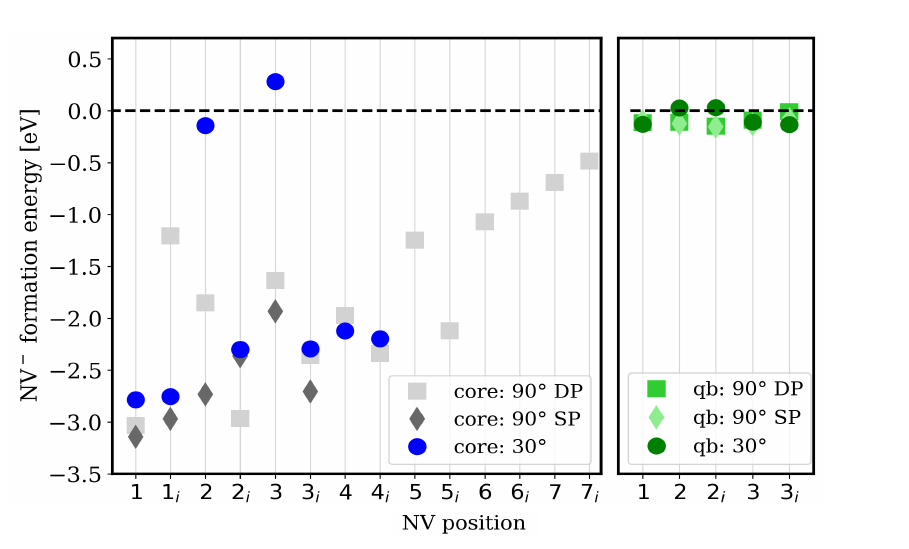}}
\caption{Triplet formation energies of the NV$^-$ defect at various core positions relative to the corresponding value in bulk diamond (cf.\ Eq.\ \ref{eq:Eform}) are plotted in the left panel in blue and gray colors for the  $30^\circ$  and  $90^\circ$ dislocations, respectively. The energies for quasi-bulk (qb) positions are plotted in the right panel in different green colors. NV positions refer to the numbering given in Fig.~\ref{fig:dis-nv}. 
}\label{fig:form_energy}
\end{figure}

First, we evaluate the formation energy of the NV$^-$ defect for those positions described in Fig.~\ref{fig:dis-nv} and constrained to the triplet electronic configuration. The results depicted in Fig.~\ref{fig:form_energy} indicate that 
most positions near the dislocation core are energetically favorable, as they gain up to $3.5$ eV energy with respect to the NV center at a bulk location. The quasi-bulk positions in the dislocation-core supercell have, as expected, an energy very similar to the NV center at a bulk location in the single-crystal supercell.

In the case of the $30^\circ$ dislocation we examined the NV$^-$ defect at all the possible configurations around the core. One can categorize the energetic stability of a NV$^-$ defect with respect to this specific dislocation core as follows:  

(i) The two positions $1$ and $1_i$, at which the symmetry axis of the NV$^-$ defect is aligned parallel 
to the dislocation line, are energetically the most stable configurations. These two configurations are structurally and electronically identical. 

(ii) Positions with a vacancy right at 
the center of the dislocation core lead to configurations in which most of the defect atoms (The N atom and the three C atoms close to the vacancy) lie in the dislocation glide plane. 
These configurations ($2_i$, 3$_i$ and $4_i$) have similar formation energies. 
 
(iii) Those NV defect positions, in which some atoms of the defect complex are not next to the geometric center of the core, have relatively higher energies. Examples are positions 2 and 3. Position 4 exceptionally gives a lower formation energy because it lies on the site of the stacking fault. 

In the case of the two $90^\circ$ dislocation cores there are many more possible positions to 
accommodate the NV defects. In order to save computational 
resources we took advantage of the above trends in energy for the $30^\circ$ dislocation core and only examined positions which 
are directly at the core or in the dislocation glide plane.
However, unlike at the $30^\circ$ dislocation core, in neither SP nor 
DP cores of $90^\circ$ dislocation there are a symmetrically meaningful (trigonally uniaxial) orientation of the NV defect parallel to the dislocation line. 

As a general trend, a point defect may have more relaxation degrees of freedom when it lies at a low-symmetry position at the center of a dislocation core. Therefore, its formation yields an energy gain and concomitantly dislocation cores can potentially trap NV centers. This may be considered an advantage or a disadvantage towards a quantum technology application. 
A potential advantage is that the dislocation core offers a possibility to assemble an array of aligned NV centers with potentially desirable properties. This could be beneficial for the design of a qubit register of a quantum computer. On the other hand, the utility of such a spin-chain system inside the structurally distorted dislocation core is under question if the NV center properties are modified too much and are no longer appropriate. To clarify this, the stability of the triplet ground state of the NV$^-$ defects at the dislocation core needs to be examined. 

\begin{figure}
{\includegraphics[width=1.\columnwidth]{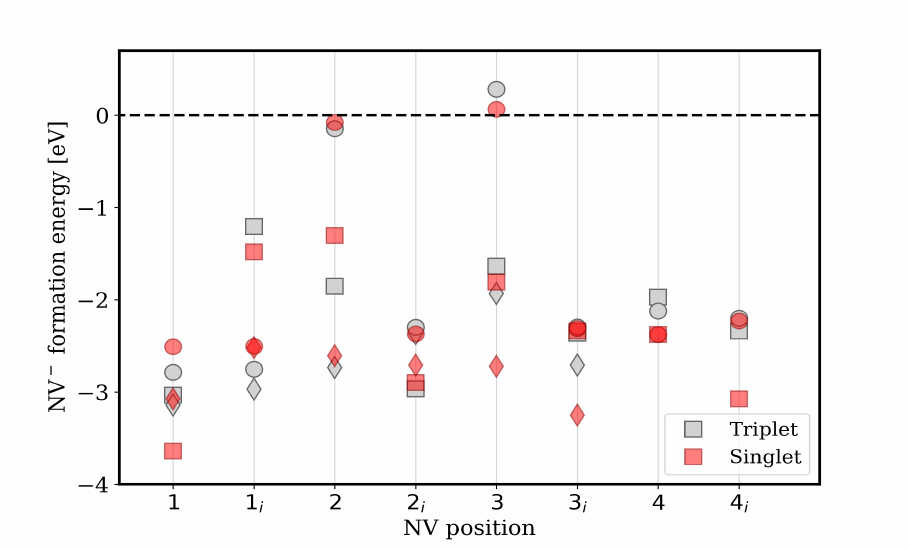}}
\caption{Comparison of formation energies of NV$^-$ defects with singlet (red) and triplet (gray) configuration, both relative to the formation energy of a triplet defect in bulk diamond (cf.\ Eq.\ \ref{eq:Eform}), for different low-energy core positions. The dislocation type is indicated by the following symbols: circle for $30^\circ$, diamond for $90^\circ$ SP and 
square for $90^\circ$ DP.  }\label{fig:energy_st}
\end{figure}

The total energy of the electronic triplet configuration of a NV$^-$ defect in a perfect single crystal is $\approx 600$ meV lower that the energy of the singlet configuration. Since this difference is much higher than thermal fluctuations at room temperature (k$_B$T$\approx 25$ meV), the stable ground state of the NV$^-$ defect in a diamond crystal is a triplet state. However, in the distorted environment of a dislocation core it can happen that the electronic singlet configuration is more stable. A comparison between formation energies of the NV$^-$ defect in a singlet or triplet configuration at various low-energy positions of the considered dislocation cores is depicted in Fig.~\ref{fig:energy_st}. 

At the $90^\circ$ SP and DP dislocation cores we do not find any NV$^-$ defect with a stable triplet configuration. Most NV$^-$ defect configurations in the $90^\circ$ DP dislocation core exhibit an electronic singlet ground state, among which the core position $1$ has the lowest formation energy with a significant energy difference of at least $\approx 500$ meV to any other positions. 
The same is observed in the case of $90^\circ$ SP dislocation core. The singlet formation energy of the NV$^-$ defect in position $3_i$ is lower by $110$ meV compared to the stable triplet ground state at the position $1$. 
The situation is different for NV $^-$ defects at a $30^\circ$ dislocation core.  
Here we found that the triplet configuration of position 
$1$ (and thus $1_i$ as well), in which the symmetry axis of NV is parallel to the dislocation line, is up to $300$ meV lower in 
energy than the singlet configuration at the same position. It is also lower than any other singlet or triplet energies of all other positions. Therefore, a $30^\circ$ partial dislocation in a diamond crystal potentially can attract NV centers while preserving their important spin triplet state. However, its features may be somewhat different from those of a NV center in a bulk crystal. In the following we will focus on an NV center at this specific favorable core position and show the modification of its electronic states, structural and magnetic properties. For completeness, appendix A provides a summary of properties of the other NV configurations that show a triplet ground state but which are energetically less likely to be observed.

\subsection{Electronic levels of NV centers}\label{sec:results:elec}

In pure diamond, far from surfaces or other structural defects, NV centers have a characteristic system of sharp electronic 
levels located in the band gap (Fig.~\ref{fig:nv})~\cite{gali77}.
The six valence electrons of the NV$^-$ defect with the $C_{3\nu}$ 
symmetry of the defect complex contribute to two fully symmetric one-electron states ($a_1$) and one doubly degenerate 
($e$) state. Two of the six electrons occupy the $a_1$ states inside the valence band (in Fig.~\ref{fig:nv}(b) it is 
marked as $a_1(1)$ and can be either $a_1(1)\uparrow$ or $a_1(1)\downarrow$ representing the spin up and down). The other four electrons contribute to levels lying in the band gap of diamond (in Fig.~\ref{fig:nv}(b) 
marked as $a_1(2)$ and $e_{x,y}$) and are well separated from the band edges~\cite{gali77,chou17b}. In the bulk interior 
the doubly degenerate states $e_{x,y}$ are mainly formed by the three equivalent C atoms next to the C vacancy. If the 
symmetry of these three carbon sites is reduced by the proximity to an extended crystal defect the degeneracy of the levels $e_x$ and $e_y$ is split. 

\begin{figure}
{\includegraphics[width=1.\columnwidth]{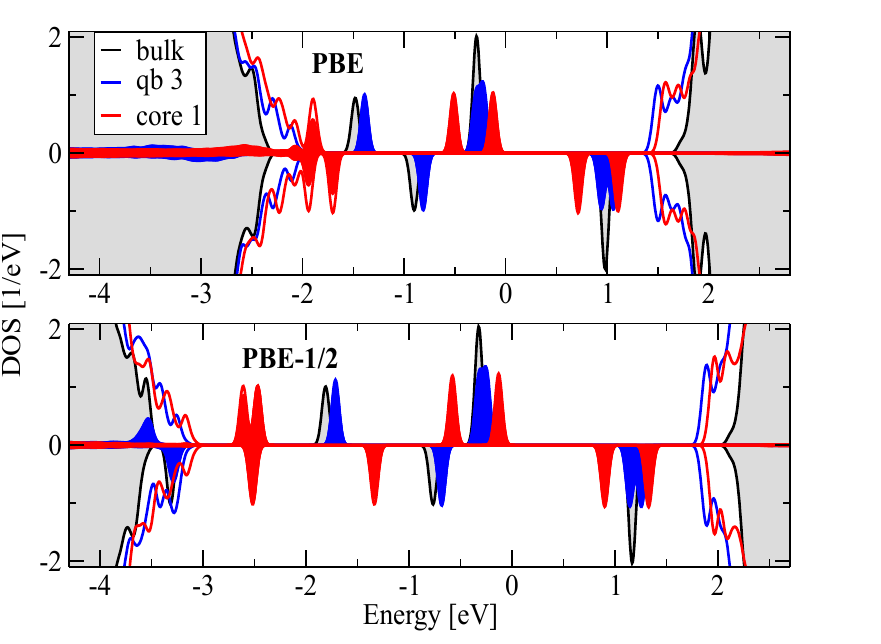}}
\caption{Total density of states (DOS) of a NV$^-$ defect in diamond placed at the following positions: quasi-bulk~(qb)~$3$ (blue) and core~$1$ (red) in the vicinity of $30^\circ$ dislocation core, and in the single crystal of diamond (gray shaded 
area); the DOS are calculated with two methods:  PBE (upper panel) and PBE-1/2 (lower panel). For the $30^\circ$ dislocation core the
contributions of the NV$^-$ defect states to the total DOS are distinguished by regions shaded in blue and red colors. All energy levels are fully
occupied up to zero energy. }\label{fig:dos}
\end{figure}

Figure~\ref{fig:dos} displays the total DOS of the diamond supercell including a NV$^-$ defect at the core position $1$ (red line) or at the quasi-bulk position $3$ (blue line) of the $30^\circ$ dislocation core. The NV$^-$  defect levels are distinguished by color-filled peaks respectively. For comparison, the total DOS of the NV$^-$ defect in the single crystal is plotted in gray shading. 

The upper panel of Fig.~\ref{fig:dos} shows the PBE calculation of the DOS. For a bulk diamond crystal including just an NV$^-$ defect (gray region), the PBE result underestimates the experimental value of $5.47$eV for the diamond band gap by $1.37$ eV. However, the NV$^-$ defect levels within the gap are well reproduced, as compared to results of more sophisticated approaches like HSE calculations~\cite{razin21v}.
For a diamond supercell including both a NV$^-$ defect and the $30^\circ$ dislocation core (blue or red line), the interaction of the defect center with the dislocation core reduces the band gap by nearly $1$ eV, as compared to the band gap of the single crystal. This is due to the formation of some extra defect levels at the valance band maximum (VBM) and conduction band minimum (CBM) which belong to the carbon atoms located inside the dislocation core. 

Using the DFT-1/2 method with PBE (lower panel of Fig.~\ref{fig:dos} labeled as PBE-1/2) the most obvious improvements visible in the DOS are (i) the diamond band gap ($\approx 5.75$ eV), which is now much closer to the 
experimental value of $5.47$ eV~\cite{wort08}, and (ii) the location of the defect levels, specifically of the $a_1(1)$ states, which lie below the VBM and have much sharper energy peaks with PBE-1/2 than with PBE. In the following we discuss the deviation of the single ($a_1$) and the double degenerate ($e_{x,y}$) defect levels of the NV center at the non-bulk environments in two separate subsections. 

\subsubsection {$a_1$ single-electron states}

The NV$^-$ defect levels of the quasi-bulk positions in the $30^\circ$ dislocation core model do not change massively compared to the bulk levels, resulting in similar DOS profiles, irrespective from using PBE or PBE-1/2 for the calculation of the DOS. Therefore, as long as the local environment of the NV center is not too much different from the one in the single crystal, PBE leads to a fair understanding of the defect levels. 
 
However, in the  core position $1$ the two spin-up levels ($a_1(1)\uparrow$, $a_1(2)\uparrow$) and the two spin-down levels ($a_1(1)\downarrow$, $a_1(2)\downarrow$) of the NV$^-$ defect (red shaded area in the upper panel of Fig.~\ref{fig:dos} with energy below $-1$ eV) are at the valence band edge, partly as the consequence of the PBE band-gap underestimation. 
Using PBE-1/2 those states are detaching from the valence band and their peaks get sharper but their positions deviate from those of the perfect bulk (or the quasi-bulk) singlet states. This deviation is observed as the blue-shift in energy of the $a_1(1)$ states from the valence band edge inside the gap and very close to the $a_1(2)$ states. Half of the contribution to the $a_1$ single electron states 
comes from the N atom and the three C atoms together contribute the other half. Therefore, changes in the spatial distribution of the electron density of defect atoms at distorted environments can change the DOS profile of these single (non-degenerate) defect levels.

\subsubsection {$e_{x,y}$ doubly degenerate electron states}
When comparing defect states of a NV center in a dislocation core position and in a bulk (or quasi-bulk) position in terms of their DOS profiles calculated with both PBE and PBE-1/2, the major difference occurs in the single electron states discussed in the previous subsection. 
In contrast the doubly degenerate $e_{x,y}$ states, which are the characteristic spin levels of a  NV center, are split at the dislocation-core position, as compared to the bulk position, without any strong energy shift or complicated mixing behavior. Looking at the DOS profiles of the $e_{x,y}$ states for the bulk, quasi-bulk and core~$1$ positions of the NV center, we see that this peak is getting broader and then splits into two separate peaks. 
In our DFT calculations with different supercell sizes and orientations of the diamond single crystal we observed that the computational error in level splitting of the $e_{x,y}$ states can rise up to maximum $10$ meV if the $C_{3\nu}$ symmetry of the NV center is not enforced in the VASP calculations. This artifact is much smaller than the significant splittings obtained for the quasi-bulk and the core position $1$, which are as large as $\approx60$ and $\approx400$ meV respectively.
Considering that the three C atoms next to the vacancy of the defect complex have the major contribution to the $e_{x,y}$ states, their deviation from perfectly degenerate bulk $e_{x,y}$ states can be explained by the changes in the spin densities of these three C atoms. Therefore, in the next section we analyze the structural distortions of the NV center at quasi-bulk or core positions compared to a bulk NV center, and we relate the structural changes to the above discussed changes in the DOS.

\subsection {Structural distortions of NV centers}\label{sec:results:geo}

\begin{figure}[]
\begin{center}
\includegraphics[width=1.\columnwidth]{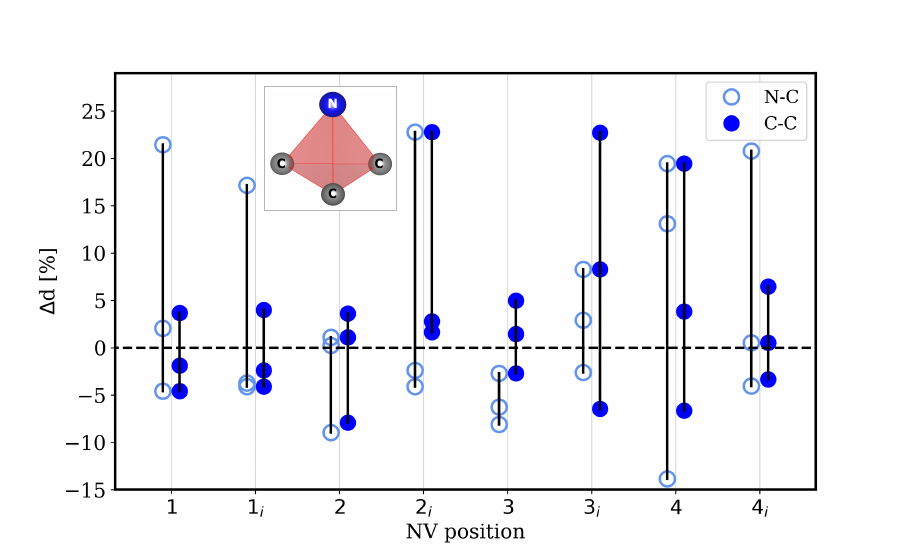}
 
\end{center}
\caption{Deviation of atomic distances $d$ between the atoms forming the tetrahedral arrangement around the NV$^-$ defect at core positions of the $30^\circ$ dislocation with respect to their bulk crystal value $d_0$. Here, $\Delta d =(d-d_{0})/d_{0}$ for C-C and C-N distances. 
Vertical solid black lines mark the deviations from an equilateral triangle of the three C atoms next to the vacancy. Schematic sketch of the tetrahedral arrangement of one N and three C atoms next to the vacancy is shown as an inset.   }\label{fig:geo-bonds}
\end{figure}

The atomic arrangement of the NV center in the bulk environment has an equilateral tetrahedral shape (see inset of Fig.~\ref{fig:geo-bonds}). For a perfectly symmetric NV center the lengths of the three N-C edges are equally $2.736~\rm{\AA}$ and the three C-C edges are equally $2.667~\rm{\AA}$. Lifting the equality of the C-C or N-C edges, when the NV center is distorted can have minor or major effects on the defect levels in the DOS. By analyzing the structural distortion of this tetrahedron in a $30^\circ$ dislocation core, we characterize the splitting observed in the $e_{x,y}$ state. Fig.~\ref{fig:geo-bonds} shows the length variation of the C-C and the N-C edges of the NV center at core positions in the $30^\circ$ dislocation supercell with respect to the bulk crystal. 
For the NV center positions in the quasi-bulk region we observe that the splitting of the $e_{x,y}$ state is less than $80$ meV
when the inequality of triangle sides is less than $2\%$ and it increases to $180$ meV if the defect deformation increases to $3\%$. Correspondingly, the NV$^-$ defects at the core positions experience larger structural distortions and therefore they have
larger level splittings in the DOS. The most favorable configuration in the dislocation core, the  core position $1$, exhibits the least 
deformation of all the core positions. The amount of this deformation is $\approx 10\%$ in the C-C distances and $\approx 25\%$ in the N-C distances. As mentioned in Sec.~\ref{sec:results:elec} the three C atoms at the C vacancy mostly contribute in the doubly degenerated $e_{x,y}$ state. Therefore, an inequality of the C-C distances has a dominant effect on the level splitting of the $e_{x,y}$ state compared to C-N distances. Respectively, in the core positions with large deformation of the NV center tetrahedron (more than $25\%$ of the C-C distances), levels are split by more than $1$ eV
(e.g., $2_i$,$3_i$ and $4$). Hence, no triplet ground state exists for such a strongly distorted NV$^-$ defect complex.

\subsection{Zero-field splitting}\label{sec:results:zfs}

The zero-field splitting originates in the spin-spin dipole interaction between unpaired electrons and it is typically analyzed in terms of an 
axial ZFS component $D$ (along the $C_{3\nu}$ symmetry axis of the NV center) and a traversal component $E$.
Felton et al.~\cite{fel09} have measured $D$ as $2.872(2)$ GHz with a high precision. 
Our theoretical values for bulk diamond exceed the measured value by about 3 to 6\% depending on the supercell size and 
other computational details. A discussion of the limitations and approximations can be found in our previous 
work~\cite{ko21} and in Ref.~[\onlinecite{bik20}] which addresses the central problem of the so-called spin contamination. 

The transversal ZFS component $E$ describes the anisotropy of the spin density between the two direction perpendicular to the axial 
component. In a perfect bulk crystal without any external magnetic field there is a perfectly symmetric 
spin distribution on the three C atoms of the NV center and consequently 
$E=0$. Crystal defects and impurities lead to stress fields and distortion in the 
crystal structure and cause a finite value for $E$. 

\begin{table}[]
	\vspace{0.2cm}
	\begin{tabular}{l c c c c c c c c}
	\hline 
   	&  & core~$1$   &  &  quasi-bulk~$3$  &  &  bulk &  &  experiment$^a$ \\
	\hline 
	$D$ [GHz]  & & 3.08 &   &  3.13  & & 2.99  & & 2.872  \\
	$E$ [MHz]  & & 120 & &  32 &  & 0   &  & 0  \\
	\hline  
	\end{tabular}
\caption{Axial ($D$) and transversal ($E$) ZFS components of the NV center in core position $1$ and quasi-bulk position $3$ in 
the $30^\circ$ partial glide dislocation supercell compared to an NV center in the pure bulk crystal of the same size supercell 
and experimental values from $^a$Ref.~\cite{fel09}. }
\label{tab:zfs}
\end{table}

Accordingly, as shown in Fig.~\ref{fig:geo-bonds}, the formation of a NV center at a dislocation core leads to an extreme 
deformation in its atomic environment and therefore can lead to deviation of ZFS parameters compared to the perfect bulk values. 

As can be seen from Tab.~\ref{tab:zfs}, the structural distortion of the NV center and its surrounding environment in the vicinity of dislocations leads to relatively 
large non-zero values of $E$. However, the $D$ component of ZFS varies only within our 
computational error ($\approx 3\%$). In fact, our ZFS analysis indicates that the $D$ value for NV center in the distorted 
environment of a dislocation is not deviating drastically compared to the NV center positioned in a bulk crystal . 
However, the opening of a spin channel (finite $E$) may change the 
NV center spin coherency features such as the coherence time T$_2$, which is interesting for applications regarding the central spin coupling to its 
environment. This has been studied by Onizhuk et al. for the case of divacancies in SiC, in which the coherence time T$_2$ of this spin defect can increase at the avoided crossing of electron spin levels at zero or low magnetic fields, where this splitting is higher than the hyperfine interaction with the nuclei spins~\cite{oniz21}. In this sense, it is also useful to check for changes in the other coupling mechanisms of a NV center with 
its surrounding atoms in a distorted environment, such as hyperfine constant parameters.   

\subsection{Hyperfine splitting of isotopes $^{14}{\rm N}$ and  $^{13}{\rm C}$ }\label{sec:results:hyperfine} 

The hyperfine interaction describes the interaction between electron spins and the nuclear spins. 
For the level spectrum of the NV$^-$ defect the nuclear spins of $^{13}{\rm 
C}$ isotopes as well as $^{14}{\rm N}$ atoms are relevant. Related level splittings of the NV$^-$ defect are of order MHz.  
 
\begin{table}[]
\begin{tabular}{l c c c c }
\hline 
   & & $A_{xx}$ \hspace{0.3cm} & $A_{yy}$  \hspace{0.3cm}  &  $A_{zz}$ \hspace{0.3cm}\\
\hline 
\multirow {4} {*} {core $1$} \hspace{0.3cm} & $^{13}{\rm C1}$\hspace{0.3cm} & 119\hspace{0.3cm} & 119\hspace{0.2cm} & 194 \hspace{0.3cm}\\
&  $^{13}{\rm C2}$\hspace{0.3cm}  & 118 \hspace{0.3cm} & 118\hspace{0.3cm}  & 201\hspace{0.3cm}\\
&  $^{13}{\rm C3}$\hspace{0.3cm}  & 127\hspace{0.3cm}  & 127\hspace{0.3cm}  & 206\hspace{0.3cm}\\
&  $^{14}{\rm N}$ \hspace{0.3cm} & -0.14 \hspace{0.3cm} & -0.13\hspace{0.3cm}  & 0.7\hspace{0.3cm}\\ \hline
               
\multirow {2} {*} {quasi-bulk $3$}  \hspace{0.3cm} & $^{13}{\rm C}$ \hspace{0.3cm} & 123\hspace{0.3cm}  & 123\hspace{0.3cm} & 204\hspace{0.3cm}\\
&  $^{14}{\rm N}$\hspace{0.3cm} & -1.9\hspace{0.3cm} &  -1.9\hspace{0.3cm}  & -1.5\hspace{0.3cm}\\
 \hline

\multirow {2} {*} {bulk}  \hspace{0.3cm} & $^{13}{\rm C}$\hspace{0.3cm} & 120\hspace{0.3cm}  & 120\hspace{0.3cm} & 200\hspace{0.3cm}\\
&  $^{14}{\rm N}$\hspace{0.3cm} & -2.1\hspace{0.3cm} &  -2.1\hspace{0.3cm} & -1.7\hspace{0.3cm}\\ \hline
\multirow {2} {*} { experiment$^a$}  \hspace{0.3cm} & $^{13}{\rm C}$\hspace{0.3cm} & 120.3\hspace{0.3cm} & 120.3\hspace{0.3cm} &  199.7\hspace{0.3cm}\\
&  $^{14}{\rm N}$\hspace{0.3cm} & -2.7\hspace{0.3cm} & -2.7\hspace{0.3cm} & -2.1 \hspace{0.3cm}\\
\hline  
\end{tabular}
\caption{ Hyperfine-structure constants, $A_{xx}$, $A_{yy}$, $A_{zz}$ in MHz, of the NV center placed in core position$1$ and quasi-bulk position$3$ in the supercell of the $30^\circ$ partial dislocation core compared to the HFS constants of the NV center of the pure bulk crystal calculated in a supercell of the same size and shape and to experimental HFS values from $^a$Ref.~\cite{fel09}. }
\label{tab:hyperfine}
\end{table}

The hyperfine-structure (HFS) constants, $A_{xx}$, $A_{yy}$ and $A_{zz}$, are listed in 
Table~\ref{tab:hyperfine} for the NV center in distorted positions compared with perfect bulk and the experimental values. 
We only list the HFS constants of the NV center defect atoms, meaning the three C atoms next to the vacancy  and the 
N atom. Since the three C atoms next to the vacancy in a perfect bulk crystal are equivalent the corresponding HFS constants are also equal. 
This equivalence in the quasi-bulk position of the NV center in a dislocation core supercell is only minorly disturbed and its effect is negligible. 
In contrast, at the  core position $1$, the HFS constants of the three C atoms are no longer equal. We label them 
separately as C1, C2 and C3, in which C1 and C2 atoms are in the same atomic layer of the dislocation glide plane and C3 is in the 
layer above. Nevertheless, the deviation of the 
HFS constants of the three $^{13}{\rm C}$ atoms compared to the bulk value is quite small (less than $3\%$, corresponding to a few MHz). 
The absolute changes of the HFS constants of the $^{14}{\rm N}$ atom are of similar magnitude. 
However, since their bulk value is of order 2 -- 3 MHz, a change of a few MHz may have a drastic effect. 
This might be especially relevant, since the nuclear spin of the $^{14}{\rm N}$ unavoidably is present in direct vicinity of the electronic spin degree of freedom of the NV$^-$ center.

\section{Summary and Conclusions}\label{sec:summary}
In this work we have investigated NV$^-$ defect complexes at cores of  $30^\circ$ and $90^\circ$ partial glide dislocations in diamond. 
We found that a dislocation core can be an attractive region for NV centers, since it provides more degrees of freedom for geometrical relaxation of the point defects. 
Based on our results, the triplet ground state configuration of NV$^-$ defect complex at the $90^\circ$ dislocation cores is not stable.  
On the contrary, there is a particular configuration of the NV center at the 
$30^\circ$ dislocation core, for which the NV center symmetry axis aligns with the dislocation line. 
This configuration not only preserves the triplet ground state but also has the lowest formation energy compared to any other NV position in both singlet or triplet ground state. 

We have analyzed some of the important features of a NV center at this 
specific dislocation core position, namely the electronic DOS, the ZFS tensor components,
and the HFS tensor components for $^{13}{\rm C}$ isotopes and $^{14}{\rm N}$ in the NV$^-$ defect complex.

The asymmetry in lateral distances of the three C defect atoms is the least in this specific core position compared to the other studied NV configurations. This deformation is still $\approx10\%$ of the prefect bulk environment which leads to a level-splitting in the doubly degenerated $e$ state and a finite value in the transversal ZFS component $E$. The axial ZFS component $D$ as well as the HFS constants related to the $^{13}{\rm C}$ have values similar to the bulk crystal (within $3\%$). 
In addition, we observed a reduction in the hyperfine constants of the $^{14}{\rm N}$ as the consequence of the elongation of the N-C distances at the core of the $30^\circ$ dislocation. 

In conclusion, the core of a $30^\circ$ partial dislocation in diamond may be a suitable location to accommodate individual NV centers because of not only preserving their spin-triplet properties, which are essential for their use in quantum sensing or quantum computing applications, but also allowing a linear-chain assembly of several NV centers, whose collective behavior may become advantageous for quantum technology. 
We hope that our theoretical study may foster further experimental and theoretical efforts to explore practical capabilities and utilities of such collective qubit arrays.

\section{Acknowledgments}
This work is funded by the Ministry of Economic Affairs, Labour and Tourism Baden W\"urttemberg in the frame of the Competence Center Quantum Computing Baden-W\"urttemberg (project "SiQuRe").
The calculations were performed on the supercomputer HOREKA funded by the German Ministry of Science, Research and the Arts Baden-W\"urttemberg and by the Federal Ministry of Education and Research.

\section{Appendix}
Here, in Tab. \ref{tab:zfs-appendix}, we briefly summarize the energetic and ZFS-parameters of NV positions at the three considered  dislocation cores which show a stable triplet ground state. However, these triplet-state formation energies are higher than that of the energetically most favorable NV configuration for the respective dislocation core so that these configurations are less likely to be observed.

\begin{table}[]
	\vspace{0.2cm}
	\begin{tabular}{l c c c c c c c c c c c c }
	\hline 
   	&  pos. && conf.  && $\Delta E_{f}$ && $\Delta E_{TS}$  && D && E \\
    &       &&        && [eV] && [eV] && [GHz] && [MHz]  \\
	\hline 
	\multirow{2}* {$30^\circ$} 
	 & 1/1$_i$ && T && -2.79 && -0.28 && 3.1 && 120   \\
	 & 2      && T && -0.14 && -0.07 && 2.9 && 328  \\
	\hline 
	\multirow{4}* {$90^\circ$ SP} 
	& 3$_i$ && S && -3.25 &&  0.79 && -   && -    \\
	& 1     && T && -3.14 && -0.07 && 2.7 && 530 \\
	& 1$_i$ && T && -2.97 && -0.45 && 2.8 && 380  \\
	& 2     && T && -2.73 && -0.13 && 3.2 && 249  \\
	\hline 
	\multirow{4}* {$90^\circ$ DP} 
	& 1     && S && -3.64 &&  0.61 && - && -       \\
	& 2     && T && -1.85 && -0.54 && 3.0 && 282  \\
	& 2$_i$ && T && -2.97 && -0.07 && 2.0 && 457  \\
	& 3$_i$ && T && -2.36 && -0.02 && 1.4 & &  3  \\
	\hline  
	\end{tabular}
\caption{Summary of properties of possible NV center positions at the three studied dislocation cores which exhibit a stable triplet (T) ground state electronic configuration.  The first row for each dislocation type corresponds to the energetically most stable NV position and configuration, which is a singlet (S) ground state in case of the two 90$^\circ$ dislocations. $\Delta E_{f}$ is the (ground state) formation energy with respect to an NV center in a bulk crystal and $\Delta E_{TS}$ is the difference between the triplet and the singlet state formation energy. Furthermore, axial ($D$) and transversal ($E$) ZFS components are given.}
\label{tab:zfs-appendix}
\end{table}

\end{document}